\documentclass[a5paper, 10pt]{article}

\usepackage{graphicx}                % for importing graphics into figures
\usepackage{float}                   % better control of figure placement
\usepackage{amsmath}                 % better mathematical typesetting
\usepackage{hyperref}                % for clickable URLs and email addresses
\usepackage[labelfont=bf]{caption}   % make caption labels boldface
\usepackage[bottom]{footmisc}		 % footnotes at bottom of page
\usepackage{mathtools}               % mathtools and amsmat
\usepackage{xurl}

\usepackage[footskip=0.5cm, headheight=1.5cm, top=2.5cm, bottom=1.5cm, left=2.1cm, right=2.1cm, a5paper]{geometry}

\usepackage{url}

\usepackage[authoryear,round,sort&compress]{natbib}

\usepackage[usenames,dvipsnames,svgnames,table]{xcolor}
\usepackage{ragged2e}
\usepackage{tabularx}
\usepackage{booktabs}
\usepackage{threeparttable} % [flushleft]
\newcolumntype{L}{>{\raggedright\arraybackslash}X}
\newcolumntype{C}{>{\centering\arraybackslash}X}
\usepackage[singlelinecheck=false,justification=centering]{caption}

\usepackage{tikz}
\usetikzlibrary{arrows, calc, matrix, patterns, positioning, shapes, trees}
\usepackage{pgfplots}
\pgfplotsset{every tick label/.append style={font=\footnotesize}}
\pgfplotsset{compat=1.17}

\title{On the role of tournament design in sporting success: A study of the North, Central American and Caribbean qualification for the 2022 FIFA World Cup}
\author{L\'aszl\'o Csat\'o  \vspace{0.25cm} \\
Institute for Computer Science and Control (SZTAKI), \\
E\"otv\"os Lor\'and Research Network (ELKH) \vspace{0.25cm} \\
Corvinus University of Budapest (BCE)}

\date{\today}

\begin{document}

\maketitle

\begin{abstract}
\noindent
Playing in the FIFA World Cup finals is an ambition shared by several nations. Since, besides luck and skill, the probability of qualification depends on the design of the qualifiers, the study of these competitions forms an integral part of sports analytics. The Confederation of North, Central America and Caribbean Association Football (CONCACAF) announced a novel qualifying format for the 2022 FIFA World Cup in July 2019. However, the COVID-19 pandemic forced the organisers to return to a more traditional structure.
The present chapter analyses how this reform impacted the chances of the national teams to qualify. It is found that the probability of participating in the FIFA World Cup finals can change by more than 5 percentage points under the assumption of fixed strengths for the teams. The idea behind the original design, to divide the contestants into two distinct sets, is worth considering due to the increased competitiveness of the matches played by the strongest and the weakest teams. We recommend mitigating the sharp nonlinearity caused by the seeding policy via a probabilistic rule to the analogy of the NBA draft lottery system.
\end{abstract}

%\keywords{OR in sports; FIFA World Cup; football; simulation; tournament design}

%\AMS{62F07, 68U20}
% Ranking and selection
% Simulation

%\JEL{C44, C63, Z20}
% Operations Research, Statistical Decision Theory
% Computational Techniques, Simulation Modeling 
% Sports Economics, General

{\fontsize{12pt}{12.0pt}\selectfont \section{Introduction} \label{Sec1}}

Association football, commonly known as football or soccer (henceforth football), is probably the most popular sport around the world: the games of the 2018 FIFA World Cup were watched by more than half of the global population \citep{FIFA2018e}. Therefore, analysing the determinants of success in football, for example, the factors on which the participation of the national teams in the FIFA World Cup finals depend, is an important topic for both academicians and football fans.

One might think that the chance of qualifying is governed by the talents of the players, the skills of the coach, or the implementation of innovative tactics---in other words, elements that are independent of the decisions made by the organiser. But this is not the case. Besides the performances of the competitors, the structural dimensions of the contest have a non-negligible effect on its expected outcome, too.
\citet{Hwang1982} and \citet{HorenRiezman1985} demonstrate the role of draws (pairings) in single-elimination tournaments.
According to \citet{KrumerLechner2017}, the winning probabilities in round-robin tournaments with sequential games between three and four players are influenced by the schedule.
\citet{LasekGagolewski2018} evaluate the formats used in the majority of European top-tier football leagues with respect to their ability to produce accurate team rankings.
\citet{Csato2021a} discusses many problems of tournament design from an axiomatic perspective.

The extensive study of tournament structures is also necessary because the rules of several high profile sporting events are regularly modified. \citet{CoronaForrestTenaWiper2019} and \citet{DagaevRudyak2019} examine the effects of a seeding system reform in the UEFA Champions League. \citet{Csato2021b} compares the four tournament formats with 24 teams that have been applied in the recent World Men's Handball Championships.

Although a great number of papers have investigated the design of the FIFA World Cup finals, especially its draw and seeding policy \citep{Jones1990, RathgeberRathgeber2007, ScarfYusof2011, Guyon2015a, Guyon2018a, LalienaLopez2019, CeaDuranGuajardoSureSiebertZamorano2020}, few research have been conducted on the World Cup continental qualifiers.
\citet{StoneRod2016} provide a non-econometric analysis of the bias in the allocation of qualifying spots among the continents.
\citet{PollardArmatas2017} assess home advantage in the group stages of qualification for the 2006, 2010, and 2014 World Cups.
\citet{DuranGuajardoSaure2017} construct schedules for the South American Qualifiers to overcome the main drawbacks of the previous approach. Their proposal was unanimously approved by the participating nations and is currently being used.
\citet{Csato2021f} studies the fairness of the 2018 FIFA World Cup qualification process.

The present chapter deals with the North, Central American and Caribbean section of the 2022 FIFA World Cup qualification. This qualifying system has been chosen because its organiser, the \emph{Confederation of North, Central America and Caribbean Association Football (CONCACAF)} has planned to fundamentally restructure the tournament format, however, the COVID-19 pandemic has forced the confederation to update the rules. Therefore, we can quantify the effects of an external shock on the contestants under the assumption that their strengths do not change at all.

Our main contribution resides in showing how tournament designs can be compared and evaluated via Monte Carlo simulations in a basic statistical model based on Elo ratings. This approach is worth adopting by any sports governing body before the modification of a tournament format in order to derive a first estimation on the effects of the suggested changes that can be useful information for the decision-makers.

The results have also some policy implications. Small differences in the FIFA World Ranking translate into high differences in the probability of qualification. Since this can be judged unfair, we propose to consider a probabilistic mechanism for the seeding to replace the current deterministic rule, similarly to the NBA draft lottery system.

The remainder of the chapter is organised as follows.
Section~\ref{Sec2} presents the two tournament formats of the CONCACAF Qualifiers for the 2022 FIFA World Cup. The methodology is detailed in Section~\ref{Sec3}, and the findings from our simulations are summarised in Section~\ref{Sec4}. Finally, Section~\ref{Sec5} concludes.

{\fontsize{12pt}{12.0pt}\selectfont \section{The CONCACAF Qualifiers for the \linebreak 2022 FIFA World Cup} \label{Sec2}}

CONCACAF announced a novel competition format for the 2022 FIFA World Cup confederation qualifiers on 10 July 2019. This aimed to make the leading CONCACAF teams more competitive on the global stage and to give emerging footballing nations the chance to pursue their dreams \citep{CONCACAF2019}.
It consists of two parts organised simultaneously:
\begin{itemize}
\item
\textbf{\href{https://en.wikipedia.org/wiki/Hexagonal_(CONCACAF)}{Hexagonal} group:} The six highest ranked member associations play a home-and-away round-robin tournament. The top three teams qualify directly for the 2022 FIFA World Cup.
\item
\textbf{Qualifier for lower-seeded teams:} The member associations ranked 7--35 (29 teams) are divided into eight groups, three groups of three teams and five groups of four teams, to play home-away round-robin matches. The group winners advance to the knockout phase, composed of the two-legged home-and-away quarterfinals, semifinals, and final.\footnote{~A two-legged contest between two teams consists of two games, with each team playing at home in one game.}
\end{itemize}
The winner of the qualifier for lower-seeded teams faces the fourth-ranked team of the Hexagonal group in a home-and-away play-off to determine the CONCACAF representative in the inter-confederation play-offs for the 2022 FIFA World Cup.
The ranking of the countries was intended to be based on the FIFA World Ranking of June 2020.

This design is called the \emph{divided format} in what follows.

However, CONCACAF confirmed on 25 June 2020 that the qualifying process will change due to the disruption caused by the COVID-19 pandemic \citep{TSN2020}.
Finally, a more traditional tournament design was chosen and published on 27 July 2020 \citep{CONCACAF2020}. This is composed of three phases as follows:
\begin{itemize}
\item
\textbf{First round:} The member associations ranked 6--35 (30 teams) are drawn into six groups of five teams each, where they play single round-robin matches (two home and two away). The group winners progress to the next stage.
\item
\textbf{Second round:} The six group winners from the first round play in three home-and-away single-elimination matchups. The winners advance to the next stage.
\item
\textbf{Final round:} The three second round winners join the five highest ranked countries to play home-and-away round-robin matches. The top three teams qualify directly for the 2022 FIFA World Cup, while the fourth-placed team represents CONCACAF in the inter-confederation play-offs for the 2022 FIFA World Cup.
\end{itemize}
The ranking of the countries is based on the FIFA World Ranking of July 2020.

The draw procedure favours higher-ranked teams \citep{FIFA2020a}.
For the draw of the first round, the 30 contestants are assigned to five pots of six teams each such that Pot $k$ consists of the teams ranked between $6k$th  and $6k+5$th. The six teams from Pot 1 ranked 6--11 are pre-seeded into groups A--F: the highest ranked team (6th) occupies position A1, the next team (7th) occupies position B1, and so on, until the lowest-ranked team from Pot 1 (11th) occupies position F1. The teams from the remaining four pots are allocated to the groups sequentially according to a random draw. The position of any team is determined by the pot from which it is drawn, for example, C4 is the participant of Group C that is drawn from Pot 4.

The schedule depends on these draw positions. In each group, the following 10 matches take place: \#1 vs.\ \#2, \#1 vs.\ \#3, \#4 vs.\ \#1, \#5 vs.\ \#1, \#2 vs.\ \#3, \#2 vs.\ \#4, \#5 vs.\ \#2, \#3 vs.\ \#4, \#3 vs.\ \#5, \#4 vs.\ \#5, where the first team plays at home. Consequently, team B2 (the only team in Group B that is drawn from Pot 2) plays against B3 and B4 at home, while it plays against B1 and B5 away. Naturally, any team plays twice at home and twice away.

In the second round, the pairings are pre-determined, too: the winner of Group A vs.\ the winner of Group F, the winner of Group B vs.\ the winner of Group E, the winner of Group C vs.\ the winner of Group D.

There are no specific draw constraints.

This design is called the \emph{undivided format} in what follows.

Therefore, in the undivided tournament format, the member associations ranked 6--11 enjoy a priority in the first round as they play against the strongest opponents (the teams drawn from Pots 2 and 3) at home. Furthermore, due to the pre-seeding of the teams from Pot 1, the team ranked 6/7/8th cannot face a team ranked higher than 11/10/9th, respectively, in the second round. 

Since the draw procedure of the qualifier for the lower-seeded teams in the divided format was never announced, a similar policy is assumed there in order to prefer higher-ranked countries:
\begin{itemize}
\item
Pot 1 contains the eight teams ranked 7--14, Pot 2 contains the eight teams ranked 15--22, Pot 3 contains the eight teams ranked 23--30, and Pot 4 contains the remaining five teams ranked 31--35.
\item
The eight teams from Pot 1 are pre-seeded into groups A--H, the highest ranked team (7th) occupies position A1, the next team (8th) occupies position B1, and so on, until the lowest-ranked team from Pot 1 (14th) occupies position H1.
\item
The eight teams drawn from Pots 2 and 3, respectively, are allocated to groups A--H in a random sequence.
\item
The five teams from Pot 4 are assigned to groups D--H sequentially as provided by a random draw.
\item
The pairings in the quarterfinals are the winner of Group A vs.\ the winner of Group H, the winner of Group B vs.\ the winner of Group G, the winner of Group C vs.\ the winner of Group F, and the winner of Group D vs.\ the winner of Group E.
\item
In the semifinals, the winner of Group A or H plays against the winner of Group D or E, and the winner of Group B or G plays against the winner of Group C or F.
\end{itemize}
Hence, although higher-ranked teams are not favoured by the schedule because the groups are home-and-away round-robin tournaments, the member associations ranked 7--9 enjoy a priority in the group stage due to competing against only two other teams instead of three (albeit the additional teams are the weakest participants). In addition, the team ranked 7/8/9/10th cannot face a team ranked higher than 14/13/12/11th, respectively, in the quarterfinals due to the pre-seeding of the teams from Pot 1. Finally, the knockout bracket implies that the team ranked 7th cannot play against a team ranked higher than the 10th in the semifinals.

{\fontsize{12pt}{12.0pt}\selectfont\section{Methodology} \label{Sec3}}

For the comparison of different tournament designs, it is almost always necessary to use Monte Carlo simulations \citep{ScarfYusofBilbao2009}. Even though some limited results can be derived from historical data, this is not possible in the case of formats that were never applied in practice.

Every simulation model is based on a prediction technique for individual games. Modelling the outcome of a football match has a long history since \citet{Maher1982} first suggested Poisson models to that end. The statistical literature often attempts to improve forecasting accuracy through sophisticated time-varying \citep{BakerMcHale2018}, Bayesian \citep{CoronaForrestTenaWiper2019}, or maximum likelihood \citep{LeyVandeWieleVanEeetvelde2019} approaches.

However, if the main aim is to understand better the properties of a tournament format and to inform decision-makers about the effects of a planned reform, it is preferred to keep the prediction model as simple as possible. An elegant method is to derive the winning probabilities exclusively from the Elo ratings of the teams \citep{VanEetveldeLey2019}. This procedure was first suggested by the Hungarian-born American physics professor \emph{\'Arp\'ad \'El\H{o}} \citep{Elo1978} to rank chess players.\footnote{~It is a relatively common mistake to call the method ELO, which seems to be an acronym.
The author of the current chapter, whose mother tongue is Hungarian, has also misunderstood the name of this system as a child: since \emph{\'el\H{o}} means \emph{living} in this language, it is natural to assume that chess players have a \emph{living} rating because it is updated regularly\dots}

Consider a match played by teams $i$ and $j$, having an Elo rating of $E_i$ and $E_j$, respectively. The winning probability of team $i$ is
\begin{equation} \label{Elo_equation}
W_{ij} = \frac{1}{1 + 10^{- \left( E_i - E_j \right) /s}},
\end{equation}
where $s$ is a scaling factor. Note that $W_{ij} + W_{ji} = 1$, thus draws are prohibited. Although drawn matches are relatively common in football, it is unlikely that allowing for this possibility would substantially modify our findings. On the other hand, accounting for draws requires at least one more arbitrary parameter.

The key question here is the reliability of the Elo ratings.
The official FIFA World Ranking uses the Elo system since August 2018 with the parameter $s=600$ \citep{FIFA2018c}. As the contestants in the undivided format of the CONCACAF Qualifiers are ranked on the basis of the FIFA World Ranking of July 2020, this will be our first set of inputs.\footnote{~The divided format was intended to use the June 2020 FIFA World Ranking \citep{CONCACAF2019}, which is the same as the July 2020 ranking because no international football matches were played in the summer of 2020.}

There is also a much older alternative ranking project called World Football Elo Ratings \citep{WorldFootballEloRatings}. Its scaling factor is $s=400$ and the rating of the home team is increased by a fixed amount of $100$. In addition, its adjustment depends on the goal difference in the game, while the FIFA version accounts neither for the margin of victory nor for home advantage.

\begin{table}[t!]
  \centering
  \caption{The strengths of the teams playing in the \\ CONCACAF Qualifiers for the 2022 FIFA World Cup}
  \label{Table1}
   
\begin{threeparttable}
    \rowcolors{1}{}{gray!20}
    \begin{footnotesize}
    \begin{tabularx}{\textwidth}{lcCC} \toprule
    Country & Abbreviation & FIFA  & Elo \\ \bottomrule
    Mexico & MEX   & 1621  & 1904 \\
    United States & USA   & 1542  & 1727 \\
    Costa Rica & CRC   & 1439  & 1653 \\
    Jamaica & JAM   & 1438  & 1585 \\
    Honduras & HON   & 1377  & 1630 \\
    El Salvador & SLV   & 1346  & 1484 \\
    Canada & CAN   & 1332  & 1600 \\
    Cura{\c c}ao & CUW   & 1313  & 1414 \\
    Panama & PAN   & 1305  & 1519 \\
    Haiti & HAI   & 1285  & 1558 \\
    Trinidad and Tobago & TRI   & 1201  & 1346 \\
    Antigua and Barbuda & ATG   & 1127  & 1158 \\
    Guatemala & GUA   & 1104  & 1483 \\
    Saint Kitts and Nevis & SKN   & 1074  & 1198 \\
    Suriname & SUR   & 1073  & 1314 \\
    Nicaragua & NCA   & 1051  & 1196 \\
    Dominican Republic & DOM   & 1019  & 1183 \\
    Grenada & GRN   & 1015  & 1213 \\
    Barbados & BRB   & 1009  & 1101 \\
    Guyana & GUY   & 988   & 1220 \\
    Saint Vincent and the Grenadines & VIN   & 986   & 1179 \\
    Bermuda & BER   & 983   & 1298 \\
    Belize & BLZ   & 974   & 1139 \\
    Saint Lucia & LCA   & 953   & 1036 \\
    Puerto Rico & PUR   & 941   & 975 \\
    Cuba  & CUB   & 936   & 1269 \\
    Montserrat & MSR   & 921   & 854 \\
    Dominica & DMA   & 919   & 1024 \\
    Cayman Islands & CAY   & 897   & 914 \\
    Bahamas & BAH   & 880   & 902 \\
    Aruba & ARU   & 867   & 892 \\
    Turks and Caicos & TCA   & 862   & 802 \\
    US Virgin Islands & VIR   & 844   & 706 \\
    British Virgin Islands & VGB   & 842   & 611 \\
    Anguilla & AIA   & 821   & 552 \\ \toprule
    \end{tabularx}
    \end{footnotesize}
\begin{tablenotes} \scriptsize
\item
The ranking is based on the FIFA World Ranking of July 2020, underlying the seeding in the qualifiers.
\item
The column FIFA shows the points according to the FIFA World Ranking of July 2020. Source: \url{https://www.fifa.com/fifa-world-ranking/ranking-table/men/rank/id12884/#CONCACAF}.
\item
The column Elo shows the World Football Elo Ratings of 16 July 2020 (the day of the July 2020 FIFA World Ranking). Source: \url{https://www.international-football.net/elo-ratings-table?year=2020&month=07&day=16&confed=CONCACAF}. This list contains six nations that do not compete in the qualifiers.
\end{tablenotes}
\end{threeparttable}
\end{table}

The strengths of the national teams are reported in Table~\ref{Table1} according to both metrics. The scaling factor of the World Football Elo Ratings is lower but its variance is higher, which has a non-negligible effect on the chances of the teams. For instance, Mexico defeats Puerto Rico with a probability of 99.16\% on the basis of the World Football Elo Ratings even if it plays away, which is reduced to 96.57\% under the FIFA World Ranking. The discrepancy can partially be assigned to the different updating principles. However, a stronger factor can be the recent introduction of the Elo system in the FIFA World Ranking: the number of matches played since the summer of 2018 is probably insufficient for the necessary adjustment of the ratings.

In both tournament formats, there are two types of matches: group matches played either home or away, and single-elimination clashes. Two-legged knockout matches are worth considering as one ``long'' game because the teams focus mostly on advancing to the next round. Therefore, we adopt the methodology of the Football Club Elo Ratings project \citep{FootballClubEloRatings}, that is, these matchups are assumed to be played at a neutral field and the difference of the Elo ratings is multiplied by the square root of 2.

We use formula \eqref{Elo_equation} as follows.
For each group match, $W_{ij}$ is calculated according to the rating method chosen. For a single-elimination matchup, $W_{ij}$ is computed analogously with the above modification. A random number $\varepsilon$ is generated between $0$ and $1$. If $\varepsilon < W_{ij}$, the winner is team $i$, otherwise, the winner is team $j$. The ranking in groups depends on the number of wins, and all ties in the number of wins are broken randomly.

In each run, all matches of both the divided and the undivided formats are simulated, as well as the teams that qualify directly and the nation going to the inter-confederation play-offs are determined. The set of matches played is recorded, too. As the CONCACAF representative in the inter-confederation play-offs for the 2022 FIFA World Cup should win a two-legged clash against a team from another continent (see \citet{Csato2021f} for the details), that place is counted as half when the probability of qualification is computed. Consequently, 3.5 qualifying slots are allocated in every iteration. The whole process is repeated 10 million times to get reasonable expected values.

We think that this relatively simple approach to model the outcomes of the games has several advantages:
\begin{itemize}
\item
The FIFA World Ranking is extensively used for seeding in football tournaments;
\item
The World Football Elo Ratings proved to have good predictive power \citep{LasekSzlavikBhulai2013} and are a competitive indicator of football performance \citep{GasquezRoyuela2016};
\item
Debates around methodological details can be avoided to a certain degree;
\item
It is easy to understand and the calculations can be replicated by any stakeholder possessing some programming skills.
\end{itemize}
The scientific literature on tournament structures often applies similar techniques \citep{Appleton1995, McGarrySchutz1997, Marchand2002, Csato2020b, Csato2021b}.
As \citet[p.~534]{Appleton1995} argues: ``\emph{since our intention is to compare tournament designs, and not to estimate the chance of the player winning any particular tournament, we may within reason take whatever model determining winners that we please}''.
However, it shall be emphasised that the following numerical results are primarily for comparative purposes.

{\fontsize{12pt}{12.0pt}\selectfont\section{Results} \label{Sec4}}

\begin{figure}[t!]
\centering

\begin{tikzpicture}
%\selectcolormodel{gray}
\begin{axis}[width = 0.96\textwidth, 
height = 0.6\textwidth,
title = {Winning probabilities from FIFA World Ranking},
title style = {font = \small},
ytick distance = 0.02,
xmajorgrids,
ymajorgrids,
%xlabel = Elo rating,
xlabel style = {font = \small},
symbolic x coords = {MEX, USA, CRC, JAM, HON, SLV, CAN, CUW, PAN, HAI, TRI, ATG, GUA, SKN, SUR, NCA, DOM, GRN, BRB, GUY, VIN, BER, BLZ, LCA, PUR, CUB, MSR, DMA, CAY, BAH, ARU, TCA, VIR, VGB, AIA},
xtick = data,
x tick label style={rotate=90,anchor=east,font=\footnotesize},
enlarge x limits = {abs = 0.25cm},
ybar stacked,
bar width = 4pt,
scaled y ticks = false,
ylabel = {Difference in the probability \\ of qualification},
ylabel style = {font = \small, align = center},
y tick label style = {/pgf/number format/.cd,fixed,precision=2},
]
% CONCACAF teams
\addplot[red,pattern = grid,pattern color = red,thick] coordinates {
(MEX,-0.0413298)
(USA,-0.053304)
(CRC,0.00160125)
(JAM,0.00265155)
(HON,0.04969455)
(SLV,-0.08418665)
(CAN,0.0430653)
(CUW,0.03099525)
(PAN,0.02734165)
(HAI,0.0201202)
(TRI,0.00433535)
(ATG,-0.00013945)
(GUA,-0.0002077)
(SKN,-0.0001573)
(SUR,-0.0000182)
(NCA,-0.00003995)
(DOM,-0.00005225)
(GRN,-0.0000648)
(BRB,-0.00006175)
(GUY,-0.00004495)
(VIN,-0.00004235)
(BER,-0.0000377)
(BLZ,-0.0000235)
(LCA,-0.00002135)
(PUR,-0.00001425)
(CUB,-0.00001335)
(MSR,-0.00001195)
(DMA,-0.0000133)
(CAY,-0.00000795)
(BAH,-0.0000053)
(ARU,-0.00000175)
(TCA,-0.00000175)
(VIR,-0.0000017)
(VGB,-0.00000115)
(AIA,-0.00000095)
};
\end{axis}
\end{tikzpicture}

\vspace{0.25cm}
\begin{tikzpicture}
%\selectcolormodel{gray}
\begin{axis}[width = 0.96\textwidth, 
height = 0.6\textwidth,
title = {Winning probabilities from World Football Elo Ratings},
title style = {font = \small},
ymin = -0.09,
ymax = 0.09,
ytick distance = 0.02,
xmajorgrids,
ymajorgrids,
%xlabel = Elo rating,
xlabel style = {font = \small},
symbolic x coords = {MEX, USA, CRC, JAM, HON, SLV, CAN, CUW, PAN, HAI, TRI, ATG, GUA, SKN, SUR, NCA, DOM, GRN, BRB, GUY, VIN, BER, BLZ, LCA, PUR, CUB, MSR, DMA, CAY, BAH, ARU, TCA, VIR, VGB, AIA},
xtick = data,
x tick label style={rotate=90,anchor=east,font=\footnotesize},
enlarge x limits = {abs = 0.25cm},
ybar stacked,
bar width = 4pt,
scaled y ticks = false,
ylabel = {Difference in the probability \\ of qualification},
ylabel style = {font = \small, align = center},
y tick label style = {/pgf/number format/.cd,fixed,precision=2},
]
% Other teams
\addplot[blue,pattern = dots,pattern color = blue,thick] coordinates {
(MEX,-0.00530485)
(USA,-0.0746579)
(CRC,-0.062862)
(JAM,0.0201959)
(HON,-0.03285955)
(SLV,0.00052355)
(CAN,0.0855888)
(CUW,0.0016313)
(PAN,0.0301924)
(HAI,0.03680905)
(TRI,0.00026525)
(ATG,-0.0000011)
(GUA,0.0006221)
(SKN,-0.00000425)
(SUR,-0.0000515)
(NCA,-0.00000385)
(DOM,-0.0000023)
(GRN,-0.00000495)
(BRB,-0.00000015)
(GUY,-0.00000645)
(VIN,-0.0000015)
(BER,-0.0000534)
(BLZ,-0.00000075)
(LCA,0)
(PUR,0)
(CUB,-0.00001385)
(MSR,0)
(DMA,0)
(CAY,0)
(BAH,0)
(ARU,0)
(TCA,0)
(VIR,0)
(VGB,0)
(AIA,0)
};
\end{axis}
\end{tikzpicture}

\captionsetup{justification=centering}
\caption[The effects of the reform]{The effects of the reform: changes in the \\ probability of qualification for the national teams \\
(See Table~\ref{Table1} for abbreviations of country names)}
%\vspace{-0.1cm}
%\begin{flushleft}
%\footnotesize{\emph{Note}: See Table~\ref{Table1} for abbreviations of country names.}
%\end{flushleft}}
\label{Fig1}

\end{figure}

%\end{document}

Figure~\ref{Fig1} outlines how the undivided tournament format of the CONCACAF Qualifiers for the 2022 FIFA World Cup has modified the probability of qualification for each national team compared to the divided format. 
According to the FIFA World Ranking, the greatest winners are Honduras---which should only enter the final round---and the countries ranked 7th or closely below (Canada, Cura{\c c}ao, Panama, Haiti) because they could not qualify directly in the divided format. On the other hand, there are three losers (Mexico, United States, Salvador): these teams should play against only five other teams in the divided format but they face seven opponents in the undivided structure. Furthermore, the 6th ranked Salvador already enters the first round, hence even its participation in the last group stage is not guaranteed.
In financial terms, one percentage point in the probability of qualification translates into at least 95 thousand USD in expected prize money as all participants of the 2018 FIFA World Cup received at least 9.5 million USD \citep{FIFA2017b}.

The consequences of the reform are somewhat different if the abilities are measured by the World Football Elo Ratings, mainly due to the different distribution of strengths and the increased variance of winning probabilities. For example, Mexico has such a high chance to qualify that even the presence of more contestants in its group cannot substantially worsen its outlook. Therefore, the chances for the next strongest teams of the United States, Costa Rica, and Honduras will decrease by a higher amount. Jamaica and Salvador have a higher probability of advancing to the play-offs in the undivided format, which balances the lower chance of direct qualification. For Canada, the 7th team in the FIFA World Ranking of July 2020, the divided format is especially disadvantageous since it closes the way of direct qualification.

\begin{figure}[t!]
\centering

\begin{tikzpicture}
%\selectcolormodel{gray}
\begin{axis}[width = 1\textwidth, 
height = 0.6\textwidth,
xmin = 1175,
ymin = -0.02,
ymax = 1.02,
xmajorgrids,
ymajorgrids,
xlabel = {Points in the FIFA World Ranking},
xlabel style = {align = center,font = \small},
%ymode = log,
%log ticks with fixed point,
scaled y ticks = false,
y tick label style = {/pgf/number format/.cd,fixed,precision=2},
ylabel = {The probability of qualification},
ylabel style = {font = \small, align = center},
]
% Old format
\addplot[red,thick,only marks,mark=diamond*,mark size=3pt] coordinates {
(1621,0.9548935)
(1542,0.8484609)
(1439,0.5233437)
(1438,0.51962305)
(1377,0.291781)
(1346,0.2019715)
(1332,0.04771005)
(1313,0.0383839)
(1305,0.0342916)
(1285,0.02532985)
(1201,0.0072878)
(1127,0.00226385)
(1104,0.0015128)
(1074,0.0008221)
(1073,0.00066245)
(1051,0.00042795)
(1019,0.00023055)
(1015,0.0002134)
(1009,0.0001846)
(988,0.00011845)
(986,0.00011225)
(983,0.0001022)
(974,0.0000775)
(953,0.00004735)
(941,0.0000358)
(936,0.0000301)
(921,0.0000222)
(919,0.0000236)
(897,0.0000135)
(880,0.00000835)
(867,0.0000047)
(862,0.00000355)
(844,0.0000026)
(842,0.000002)
(821,0.0000013)
};
% New format
\addplot[blue,thick,only marks,mark=star,mark size=3pt] coordinates {
(1621,0.9135637)
(1542,0.7951569)
(1439,0.52494495)
(1438,0.5222746)
(1377,0.34147555)
(1346,0.11778485)
(1332,0.09077535)
(1313,0.06937915)
(1305,0.06163325)
(1285,0.04545005)
(1201,0.01162315)
(1127,0.0021244)
(1104,0.0013051)
(1074,0.0006648)
(1073,0.00064425)
(1051,0.000388)
(1019,0.0001783)
(1015,0.0001486)
(1009,0.00012285)
(988,0.0000735)
(986,0.0000699)
(983,0.0000645)
(974,0.000054)
(953,0.000026)
(941,0.00002155)
(936,0.00001675)
(921,0.00001025)
(919,0.0000103)
(897,0.00000555)
(880,0.00000305)
(867,0.00000295)
(862,0.0000018)
(844,0.0000009)
(842,0.00000085)
(821,0.00000035)
};
\end{axis}
\end{tikzpicture}

\vspace{0.25cm}
\begin{tikzpicture}
%\selectcolormodel{gray}
\begin{axis}[width = 1\textwidth, 
height = 0.6\textwidth,
xmin = 1455,
ymin = -0.02,
ymax = 1.02,
xmajorgrids,
ymajorgrids,
xlabel = {Points in the World Football Elo Ratings},
xlabel style = {align = center,font = \small},
%xmode = log,
%log ticks with fixed point,
scaled y ticks = false,
y tick label style = {/pgf/number format/.cd,fixed,precision=2},
ylabel = {The probability of qualification},
ylabel style = {font = \small, align = center},
legend entries = {Divided tournament format$\quad$,Undivided tournament format},
legend style = {at = {(0.5,-0.25)},anchor = north,legend columns = 1,font = \small}
]
% Old format
\addplot[red,thick,only marks,mark=diamond*,mark size=3pt] coordinates {
(1904,0.9991368)
(1727,0.89764465)
(1653,0.6268092)
(1585,0.26396225)
(1630,0.4972)
(1484,0.03259755)
(1600,0.09155615)
(1414,0.00241285)
(1519,0.03078675)
(1558,0.0447835)
(1346,0.00033385)
(1158,0.00000135)
(1483,0.01247505)
(1198,0.00000505)
(1314,0.00014355)
(1196,0.0000047)
(1183,0.00000285)
(1213,0.0000068)
(1101,0.0000002)
(1220,0.00000935)
(1179,0.00000185)
(1298,0.0000949)
(1139,0.00000075)
(1036,0)
(975,0)
(1269,0.00003005)
(854,0)
(1024,0)
(914,0)
(902,0)
(892,0)
(802,0)
(706,0)
(611,0)
(552,0)
};
% New format
\addplot[blue,thick,only marks,mark=star,mark size=3pt] coordinates {
(1904,0.99383195)
(1727,0.82298675)
(1653,0.5639472)
(1585,0.28415815)
(1630,0.46434045)
(1484,0.0331211)
(1600,0.17714495)
(1414,0.00404415)
(1519,0.06097915)
(1558,0.08159255)
(1346,0.0005991)
(1158,0.00000025)
(1483,0.01309715)
(1198,0.0000008)
(1314,0.00009205)
(1196,0.00000085)
(1183,0.00000055)
(1213,0.00000185)
(1101,0.00000005)
(1220,0.0000029)
(1179,0.00000035)
(1298,0.0000415)
(1139,0)
(1036,0)
(975,0)
(1269,0.0000162)
(854,0)
(1024,0)
(914,0)
(902,0)
(892,0)
(802,0)
(706,0)
(611,0)
(552,0)
};
\end{axis}
\end{tikzpicture}

\captionsetup{justification=centering}
\caption[The effects of the reform]{The effects of the reform: the probability \\ of qualification as the function of teams' strength \\
(Only the 11 strongest national teams are shown in both cases)}
%\vspace{-0.1cm}
%\begin{flushleft}
%\footnotesize{\emph{Note}: Only the 11 strongest national teams are shown in both cases.}
%\end{flushleft}}
\label{Fig2}

\end{figure}

%\end{document}

Figure~\ref{Fig2} focuses on the connection between strength and the probability of qualification for both tournament formats. The relationship is monotonic, except for Jamaica and Canada using the World Football Elo Ratings: the latter team has a more difficult path to qualify as the seeding is based on the FIFA World Ranking. While this is not an inherent failure of the tournament designs, the distortion is clearly lower under the undivided format where the teams are not divided preliminarily into two distinct sets.

The impact of the tournament structure cannot be neglected. For instance, the reform almost halves the probability of qualification for Salvador (from 20.2\% to 11.78\%) but it is roughly doubled for Canada (9.08\% instead of 4.78\%) if the FIFA World Ranking is used to measure the strengths of the teams. The undivided tournament design prefers most teams at the expense of some leading CONCACAF nations, which might explain the ditching of the original plan.

\begin{figure}[t!]
\centering

\begin{tikzpicture}
%\selectcolormodel{gray}
\begin{axis}[width = 0.97\textwidth, 
height = 0.6\textwidth,
xmajorgrids,
ymajorgrids,
xlabel = {Points in the FIFA World Ranking},
xlabel style = {align = center,font = \small},
%ymode = log,
%log ticks with fixed point,
scaled y ticks = false,
y tick label style = {/pgf/number format/.cd,fixed,precision=2},
ylabel = {The average strength of opponents},
ylabel style = {font = \small, align = center},
]
\draw[dashed](axis cs:\pgfkeysvalueof{/pgfplots/ymin},\pgfkeysvalueof{/pgfplots/ymin})  -- (axis cs:\pgfkeysvalueof{/pgfplots/ymax},\pgfkeysvalueof{/pgfplots/ymax});
% Old format
\addplot[red,thick,only marks,mark=diamond*,mark size=3pt] coordinates {
(1621,1427.13894661442)
(1542,1440.51213409848)
(1439,1456.49870232228)
(1438,1456.67979142352)
(1377,1469.3161267368)
(1346,1476.69647598132)
(1332,1074.65757470199)
(1313,1072.67526451627)
(1305,1073.17011766505)
(1285,1029.37160319234)
(1201,1013.85118240851)
(1127,996.17305650211)
(1104,991.174980640344)
(1074,985.857137742646)
(1073,1043.06020542233)
(1051,1040.26567519297)
(1019,1036.54633814002)
(1015,1036.05567630837)
(1009,1035.50139018794)
(988,1033.41391362612)
(986,1033.21117002516)
(983,1032.87963075338)
(974,1061.62871918097)
(953,1060.28972233864)
(941,1059.61307602592)
(936,1059.32367898419)
(921,1058.55396564244)
(919,1058.49085249299)
(897,1057.45584547653)
(880,1056.7754033669)
(867,1038.81710614527)
(862,1038.60694429228)
(844,1037.89433257657)
(842,1037.80234667034)
(821,1037.12103498341)
};
% New format
\addplot[blue,thick,only marks,mark=star,mark size=3pt] coordinates {
(1621,1356.05873562857)
(1542,1367.34444991429)
(1439,1382.05873562857)
(1438,1382.20159277143)
(1377,1390.91587848571)
(1346,1254.5854476809)
(1332,1247.53630150207)
(1313,1241.46002922082)
(1305,1238.54794065603)
(1285,1230.8461661022)
(1201,1199.29946253996)
(1127,1089.84623346465)
(1104,1077.16553752285)
(1074,1060.96531304141)
(1073,1060.3063313079)
(1051,1049.03141896061)
(1019,1033.54598343653)
(1015,968.691015132242)
(1009,965.689157450499)
(988,955.533976347773)
(986,954.743922187864)
(983,953.346598410178)
(974,949.44528771715)
(953,1102.4194912703)
(941,1099.8461088687)
(936,1098.81576561997)
(921,1095.94863890646)
(919,1095.61619905407)
(897,1091.89972212179)
(880,1193.17501644207)
(867,1192.22820944549)
(862,1191.88903948095)
(844,1190.823653504)
(842,1190.67981630532)
(821,1189.7158322967)
};
\end{axis}
\end{tikzpicture}

\vspace{0.25cm}
\begin{tikzpicture}
%\selectcolormodel{gray}
\begin{axis}[width = 0.97\textwidth, 
height = 0.6\textwidth,
xmajorgrids,
ymajorgrids,
xlabel = {Points in the World Football Elo Ratings},
xlabel style = {align = center,font = \small},
%xmode = log,
%log ticks with fixed point,
scaled y ticks = false,
y tick label style = {/pgf/number format/.cd,fixed,precision=2},
ylabel = {The average strength of opponents},
ylabel style = {font = \small, align = center},
legend entries = {Divided tournament format$\quad$,Undivided tournament format},
legend style = {at = {(0.5,-0.25)},anchor = north,legend columns = 1,font = \small}
]
\draw[dashed](axis cs:\pgfkeysvalueof{/pgfplots/ymin},\pgfkeysvalueof{/pgfplots/ymin})  -- (axis cs:\pgfkeysvalueof{/pgfplots/ymax},\pgfkeysvalueof{/pgfplots/ymax});
% Old format
\addplot[red,thick,only marks,mark=diamond*,mark size=3pt] coordinates {
(1904,1428.36131991076)
(1727,1441.07319083928)
(1653,1455.78305605563)
(1585,1456.420012367)
(1630,1466.07975276146)
(1484,1480.99592444552)
(1600,1104.07739385699)
(1414,1044.17678531485)
(1519,1080.68504281263)
(1558,1062.38854049743)
(1346,1003.18188132909)
(1158,966.44719747752)
(1483,1055.60676546265)
(1198,978.043202476887)
(1314,1050.81827812666)
(1196,1036.46698724299)
(1183,1035.22159103913)
(1213,1038.24794586343)
(1101,1028.67589972109)
(1220,1039.06536764602)
(1179,1034.94298083376)
(1298,1048.91653361607)
(1139,1059.43961803802)
(1036,1055.2081944811)
(975,1053.88752002325)
(1269,1070.15377858549)
(854,1052.7913739064)
(1024,1054.89902241174)
(914,1053.17801176589)
(902,1053.04779110434)
(892,1034.23342477278)
(802,1033.89426680686)
(706,1033.79840914645)
(611,1033.7708339486)
(552,1033.78221582062)
};
% New format
\addplot[blue,thick,only marks,mark=star,mark size=3pt] coordinates {
(1904,1372.97877202857)
(1727,1384.26448631429)
(1653,1398.97877202857)
(1585,1399.12162917143)
(1630,1407.83591488571)
(1484,1277.27481327171)
(1600,1280.52742907567)
(1414,1221.26773985753)
(1519,1286.21765102937)
(1558,1261.41732291496)
(1346,1200.47207206241)
(1158,987.664932400463)
(1483,1169.75603903733)
(1198,999.373359037236)
(1314,1053.74191376637)
(1196,998.767460403877)
(1183,994.564426796357)
(1213,926.972257043394)
(1101,902.511111450884)
(1220,929.547109633605)
(1179,917.006646147129)
(1298,966.2246720727)
(1139,908.284522139298)
(1036,1077.74542525764)
(975,1076.10110373444)
(1269,1119.21485147444)
(854,1075.09967117499)
(1024,1077.34524878942)
(914,1075.39643935525)
(902,1186.0007279609)
(892,1185.97167527087)
(802,1185.8592224987)
(706,1185.85314897598)
(611,1185.82757538307)
(552,1185.83593373754)
};
\end{axis}
\end{tikzpicture}

\captionsetup{justification=centering}
\caption[The effects of the reform]{The effects of the reform: the average \\ strength of opponents as the function of own rating \\
(The black dashed line represents the 45-degree line)}
\label{Fig3}

\end{figure}

%\end{document}

Another important attribute of any tournament is the competitiveness of the matches played, quantified in Figure~\ref{Fig3} through the average strength of the opponents. Note that in a round-robin group, this value is the highest for the weakest team. The divided tournament format essentially consists of two parallel contests, thus the six best teams of the Hexagonal group play against much stronger competitors than the lower-seeded teams, among which our metric remains approximately constant. Therefore, the weakest teams can expect more success in the divided format but the matches of the middle teams are more balanced under the undivided design. Finally, the leading CONCACAF nations should contest more competitive games on average in the divided format, too, since the Hexagonal contains fewer teams than the group in the last round of the undivided structure.

Theoretical models are also important to examine because the actual distribution of teams' strengths may hide certain features of tournament designs. To that end, we set the Elo rating of the $i$th ranked team at $1{,}300 + (18-i) \Delta$---that is, the 18th country at the middle of the ranking has a fixed strength of $1{,}300$---with a parameter $\Delta \geq 0$, and derive the winning probabilities from formula \eqref{Elo_equation} such that $s=400$ and home advantage is 100, analogously to the World Football Elo Ratings.

\begin{figure}[t!]
\centering

\begin{tikzpicture}
%\selectcolormodel{gray}
\begin{axis}[width = 1\textwidth, 
height = 0.6\textwidth,
title = {Divided tournament format},
title style = {align = center,font = \small},
xmin = 1,
xmax = 10,
ymin = -0.02,
ymax = 1.02,
ymajorgrids,
xlabel = Team rank,
xlabel style = {font =\small},
]
% Difference 0
\addplot[red,thick,mark=diamond*, mark size=3pt] coordinates {
(1,0.5415727)
(2,0.54175865)
(3,0.5416901)
(4,0.5416967)
(5,0.54175375)
(6,0.54156595)
(7,0.0104124)
(8,0.0103908)
(9,0.010432)
(10,0.0077938)
(11,0.00779535)
(12,0.007814)
(13,0.0078362)
(14,0.00778315)
(15,0.008775)
(16,0.0087978)
(17,0.0087505)
(18,0.00880535)
(19,0.008768)
(20,0.00875825)
(21,0.0087854)
(22,0.00880645)
(23,0.0088426)
(24,0.0087977)
(25,0.0087833)
(26,0.00880245)
(27,0.0088364)
(28,0.00878865)
(29,0.00877615)
(30,0.00875685)
(31,0.0078013)
(32,0.00782005)
(33,0.007818)
(34,0.00783015)
(35,0.0078041)
};

\addplot[blue,thick,mark=star, mark size=2pt] coordinates {
(1,0.8129549)
(2,0.72872745)
(3,0.6254236)
(4,0.5084028)
(5,0.39125545)
(6,0.285792)
(7,0.04958645)
(8,0.03448205)
(9,0.0230983)
(10,0.01550405)
(11,0.0099003)
(12,0.0063241)
(13,0.00390815)
(14,0.00232605)
(15,0.00101215)
(16,0.00057985)
(17,0.00033655)
(18,0.00018235)
(19,0.0001009)
(20,0.0000522)
(21,0.00002795)
(22,0.00001375)
(23,0.00000435)
(24,0.00000245)
(25,0.0000012)
(26,0.0000004)
(27,0.00000015)
(28,0.00000005)
(29,0)
(30,0)
(31,0.00000005)
(32,0)
(33,0)
(34,0)
(35,0)
};

\addplot[ForestGreen,thick,mark=otimes*, mark size=2pt] coordinates {
(1,0.94349425)
(2,0.8613002)
(3,0.70845815)
(4,0.48446435)
(5,0.26897305)
(6,0.1258809)
(7,0.05779085)
(8,0.0286989)
(9,0.0127366)
(10,0.0051589)
(11,0.00195435)
(12,0.00072145)
(13,0.00025775)
(14,0.00008635)
(15,0.0000168)
(16,0.0000053)
(17,0.00000125)
(18,0.0000004)
(19,0.00000015)
(20,0.00000005)
(21,0)
(22,0)
(23,0)
(24,0)
(25,0)
(26,0)
(27,0)
(28,0)
(29,0)
(30,0)
(31,0)
(32,0)
(33,0)
(34,0)
(35,0)
};
\end{axis}
\end{tikzpicture}

\vspace{0.25cm}
\begin{tikzpicture}
%\selectcolormodel{gray}
\begin{axis}[width = 1\textwidth, 
height = 0.6\textwidth,
title = {Undivided tournament format},
title style = {align = center,font = \small},
xmin = 1,
xmax = 10,
ymin = -0.02,
ymax = 1.02,
ymajorgrids,
xlabel = Team rank,
xlabel style = {font =\small},
legend entries = {$\Delta = 0 \qquad$,$\Delta = 20 \qquad$,$\Delta = 40$},
legend style = {at = {(0.5,-0.2)},anchor = north,legend columns = 4,font = \small}
]
% Difference 0
\addplot[red,thick,mark=diamond*, mark size=3pt] coordinates {
(1,0.4374273)
(2,0.4375556)
(3,0.4373942)
(4,0.4374999)
(5,0.4376159)
(6,0.0437126)
(7,0.04382775)
(8,0.04373365)
(9,0.0438684)
(10,0.04370165)
(11,0.04363845)
(12,0.04365115)
(13,0.04366975)
(14,0.0437925)
(15,0.0438058)
(16,0.0437596)
(17,0.04376745)
(18,0.04381525)
(19,0.04382545)
(20,0.0438193)
(21,0.04375085)
(22,0.04385415)
(23,0.04371115)
(24,0.04383755)
(25,0.0437959)
(26,0.0438547)
(27,0.0436563)
(28,0.0437333)
(29,0.0436345)
(30,0.04384355)
(31,0.0436842)
(32,0.04366175)
(33,0.0437294)
(34,0.04365935)
(35,0.0437117)
};

\addplot[blue,thick,mark=star, mark size=2pt] coordinates {
(1,0.757996)
(2,0.68623795)
(3,0.6054027)
(4,0.51806905)
(5,0.42987365)
(6,0.1758906)
(7,0.12267165)
(8,0.08212215)
(9,0.052756)
(10,0.0324885)
(11,0.01932345)
(12,0.00772415)
(13,0.0044018)
(14,0.002451)
(15,0.00134145)
(16,0.0006838)
(17,0.00035695)
(18,0.00011075)
(19,0.00005475)
(20,0.0000243)
(21,0.000011)
(22,0.0000051)
(23,0.0000018)
(24,0.0000008)
(25,0.0000004)
(26,0.0000002)
(27,0)
(28,0)
(29,0.00000005)
(30,0)
(31,0)
(32,0)
(33,0)
(34,0)
(35,0)
};

\addplot[ForestGreen,thick,mark=otimes*, mark size=2pt] coordinates {
(1,0.89826285)
(2,0.8106304)
(3,0.6793792)
(4,0.51083045)
(5,0.3393741)
(6,0.1506774)
(7,0.07007835)
(8,0.0276863)
(9,0.0093654)
(10,0.0028068)
(11,0.0007784)
(12,0.00009895)
(13,0.0000233)
(14,0.00000645)
(15,0.00000135)
(16,0.00000025)
(17,0.00000005)
(18,0)
(19,0)
(20,0)
(21,0)
(22,0)
(23,0)
(24,0)
(25,0)
(26,0)
(27,0)
(28,0)
(29,0)
(30,0)
(31,0)
(32,0)
(33,0)
(34,0)
(35,0)
};
\end{axis}
\end{tikzpicture}

\caption{Theoretical model: the probability of qualification \\
(Only the 10 highest ranked teams are shown in both cases) \\
(The lines serve for illustrative purposes)}
%\vspace{-0.1cm}
%\begin{flushleft}
%\footnotesize{\emph{Note}: Only the 10 highest ranked teams are shown in both cases. \\
%The meaning of the parameter $\Delta$ is explained in the main text. \\
%The lines serve for illustrative purposes.}
%\end{flushleft}}
\label{Fig4}

\end{figure}
%\end{document}

The chances of qualification are presented in Figure~\ref{Fig4}. $\Delta = 0$ corresponds to the fully competitive scenario, while $\Delta = 20$  and, especially, $\Delta = 40$ represent unbalanced contests. The findings convey two basic messages.
First, the sharp distinction of the teams in the divided format does not lead to misaligned incentives, that is, no team is interested in being ranked 7th to play among the lower-seeded teams instead of the Hexagonal group. Therefore, the divided tournament design is strategy-proof with respect to the initial ranking of the teams. As illustrated by the UEFA Euro 2020 qualifying tournament \citep[Chapter~6.4]{Csato2021f}, this is far from guaranteed since the 7th ranked team can obtain the CONCACAF slot in the inter-confederation play-offs for the 2022 FIFA World Cup after playing only one two-legged clash with a higher-ranked team.

Second, the probability of qualification exhibits a sharp nonlinearity, see the breaking point between the 6th and the 7th teams in the old, as well as between the 5th and 6th teams in the undivided design. While this policy might intensify the chase for a particular position in the FIFA World Ranking, it could be unfair when small differences have such powerful implications. Thus it is worth considering a lottery system: the seeding of the teams can be determined by the FIFA World Ranking according to a probabilistic mechanism instead of the current rule. For instance, the fifth team that enters only the final group stage in the undivided tournament format should not necessarily be the 5th ranked one but the 5th ranked with a probability of 60\%, the 6th with a probability of 30\%, or the 7th with a probability of 10\%. A similar procedure is used in the \href{https://en.wikipedia.org/wiki/NBA_draft_lottery}{NBA draft lottery}. Such a policy would also be able to mitigate the effects of changing from the divided to the undivided format as the comparison of Figures~\ref{Fig1} and \ref{Fig4} highlights.

{\fontsize{12pt}{12.0pt}\selectfont\section{Conclusions} \label{Sec5}}

The current chapter has aimed to assess how a reform---induced by the calendar disruption due to the COVID-19 global pandemic---in the format of the CONCACAF Qualifiers for the 2022 FIFA World Cup impacted the chances of the national teams to qualify. It has turned out that the tournament design on its own can have quite a substantial role in the probability of playing in the FIFA World Cup finals, even amounting to 5-6 percentage points under the assumption of fixed Elo ratings for the teams. Since the initial division of the contestants into two sets in the original structure of the qualifiers has increased the competitiveness of the matches played by the strongest and the weakest teams without creating misaligned incentives for being seeded lower, this solution offers a reasonable alternative to the traditional hybrid format consisting of subsequent group and knockout stages. Finally, the sharp non-linearity caused by the deterministic seeding policy is recommended to be reduced via a probabilistic rule to the analogy of the NBA draft lottery.

These proposals can contribute to making qualifying tournaments more exciting and fair. Hopefully, our chapter will also call the attention of sports governing bodies worldwide that even basic statistical models, based on widely used metrics such as the FIFA World Ranking, might give useful insights into the problem of designing a tournament.

{\fontsize{12pt}{12.0pt}\selectfont\section*{Acknowledgements}}

\noindent
We are grateful to \emph{Yves Dominicy} and \emph{Christophe Ley} for their beneficial remarks. \\
This project could not have been implemented without \emph{my father} (also called \emph{L\'aszl\'o Csat\'o}), who has coded the simulations in Python. \\
We are indebted to the \href{https://en.wikipedia.org/wiki/Wikipedia_community}{Wikipedia community} for collecting and structuring valuable information on the sports tournaments discussed. \\
The research was supported by the MTA Premium Postdoctoral Research Program grant PPD2019-9/2019.

\clearpage

\bibliographystyle{apalike}
\bibliography{All_references}

\end{document}